\documentclass[12pt]{article}
\def\beq{\begin{equation}}
\def\eeq{\end{equation}}
\def\barr{\begin{array}}
\def\earr{\end{array}}
\def\dis{\displaystyle}
\begin{document}
\thispagestyle{empty}

\begin{center}

{\Large\bf On the origin of brane cosmological constant in two-brane warped geometry model}\\ [18mm]

Sayantani Lahiri\footnote{E-mail: sayantani.lahiri@gmail.com}\\
{\em Relativity and Cosmology Centre 
\\ Department of Physics}
\\{\em Jadavpur University
\\ Raja Subodh Chandra Mullick Road
\\Jadavpur
\\Kolkata- 700032, India}  \\ [8mm]

Soumitra SenGupta\footnote{E-mail: tpssg@iacs.res.in} 
\\{\em Department of Theoretical Physics\\ Indian Association for the
Cultivation of Science
\\ 2A and 2B Raja S.C. Mallick Road, Jadavpur
\\Kolkata - 700 032, India} \\

\end{center}

\bigskip
\abstract{In the backdrop of generalised Randall-Sundrum braneworld scenario, we look for the possible 
origin of an effective four dimensional cosmological constant ($\Omega_{vis}$) on the visible 3-brane due to the
effects of 
bulk curvature and the modulus field that can either be  a constant or dependent on  extra dimensional co-ordinate $y$ or  a time dependent quantity. 
In case of constant or $y$ dependent modulus field, the induced $\Omega_{vis}$ leads to an exponentially expanding Universe. 
For such modulus fields the presence of vacuum energy densities on either of the $3$-branes as well as a non-vanishing 
bulk curvature $l$ ($l \sim {\Lambda_5}^{-1}$) are essential to generate an effective $\Omega_{vis}$. 
In particular for constant modulus field the Hubble constant turns out to be equal to the visible brane cosmological constant which agrees 
with the present result. In an alternative scenario, a time dependent modulus field is found to be capable of accelerating the Universe. 
The Hubble parameter in this case is determined for a slowly time-varying modulus field.}

\newpage
\section{Introduction}
\par The introduction of extra dimensions to four dimensional spacetime dates back to early twentieth century when Kaluza and Klein tried to unify 
gravity and electromagnetic theory in the presence of an extra spatial dimension. 
In recent times with the advent of String theory this concept once again witnessed revival. 
Moreover, our inability to explain many problems in Particle physics and Cosmology has stimulated widespread interests for exploring higher 
dimensional theories in recent times. To start with, the Standard model of Particle physics suffers from a discrepancy known as Higgs mass hierarchy 
problem or alternatively gauge hierarchy problem \cite{Dress}. 
The Higgs boson receives quantum corrections from higher order self energy corrections typically of the order of Planck energy scale, but to obtain a 
theoretical predicted value of Higgs mass (of the order of 100 GeV), an extreme fine tuning is necessary, often called the naturalness problem.  
One therefore needs to look beyond the Standard model to seek a possible explanation to this fine tuning condition. 
There also exists another fine-tuning known as cosmological constant problem which relates with the mismatch between observed and theoretically 
estimated  values of cosmological constant and needs to be resolved. 
Extra dimensional theories offer possible solutions to these problems and address many more which include various phenomenological and cosmological
issues including the existence of dark energy, inflation, CMB anisotropies etc. 
\par Typically in a braneworld model, the branes are hypersurfaces embedded in a higher dimensional bulk. All standard model particles are confined 
to the branes while gravity is allowed to propagate everywhere including in higher dimensions. 
Among various extra dimensional models, the warped geometry model assumes that the extra dimensions are compactified on a closed path 
and their sizes are required to be extremely small (Planck length $\sim 10^{-33}$ cm) such that no intermediate scale enters in the theory.
On the contrary the large extra dimensional models like ADD model \cite{arkani} tries to solve the naturalness problem by lowering the 
higher dimensional Planck scale to TeV scale with the help of two flat and compact extra dimensions whose sizes are quite large compared to the 
Planck length. 
\par In this work we focus our attention to the well known warped geometry model proposed by Randall and Sundrum (RS) \cite{randall} and it's subsequent 
generalization to include non-flat 3-branes \cite{cosmo_con}. The Randall Sundrum model consisting of two $Z_2$ symmetric $3$-branes embedded in a 
five dimensional AdS bulk offers a possible solution to the gauge hierarchy problem by exponentially suppressing mass of Higgs boson and it's vev on 
TeV brane. The naturalness problem can be avoided in this scenario without invoking any intermediate scale in the theory. 
In this model, the effect of bulk cosmological constant is exactly counterbalanced by the brane tensions leading to zero brane cosmological constant 
and therefore the 3-branes are flat and static \cite{randall}. So far various applications of RS model in the cosmology/phenomenological issues of 
Particle physics and String theory have been explored \cite{phen,lang}.
\par The flat brane RS model described above was generalised \cite{cosmo_con} to include $3$-branes which possess a net induced non-zero 
cosmological constant. We try to investigate the possible origin of this $3$-brane cosmological constant by analysing the effective gravity models 
derived by embedding $3$-branes at orbifold fixed points in a bulk spacetime \cite{shiromizu_2} in presence of bulk cosmological constant and 
a modulus field which can be  constant, time dependent or extra dimensional co-ordinate dependent.\\
In section {\ref{sec_1}} and  in section {\ref{sec_2}}, we briefly discuss the works by Shiromizu et.al \cite{shiromizu_2} and the generalised 
Randall Sundrum model \cite{cosmo_con}. In section {\ref{sec_3}}, the induced cosmological constant in the presence of a constant radion field, 
leading to an exponentially expanding Universe, has been shown to emerge due to the presence of both brane matter as well as bulk curvature. 
The Hubble parameter on the visible brane is determined in this case. 
It is shown to be proportional to the induced visible brane cosmological constant which agrees with the present results without any fine tuning. 
In section {\ref{sec_4}}, we incorporate extra-dimensional coordinate ($y$) dependence in the modulus field called radion and once again try to generate 
an effective cosmological constant on the Universe located on the visible brane ($y = r \pi$). An exponential solution of the scale factor is also 
found in this case and we have further examined the origin of visible brane cosmological constant. It is to be noted that many results of these 
two sections are similar in nature. Finally in section {\ref{sec_5}}, we find that a time dependent radion field can be responsible for an 
accelerated expanding phase  of our Universe even in absence of any brane matter. For a slowly time varying radion field, we  
determine the exact solution of Hubble's expansion of the Universe. Such a scenario indicates that a slowly time evolving inter-brane 
separation (modulus) may result into an effective accelerating scale factor of our 3-brane Universe.\\

\section{Low energy effective Einstein's equation in the presence of two branes}  \label{sec_1}
In this section we review the analysis done in \cite{shiromizu_2}.
Let us consider a system of two $3$-branes placed at the orbifold fixed points and embedded in the bulk which is a five dimensional AdS spacetime 
containing the bulk cosmological constant $\Lambda_5$ only \cite{shiromizu_2}. Each of the $3$-branes is $Z_2$ symmetric. The $3$-brane located 
at $y = 0 $ hypersurface has positive tension ${\cal V}_{pl} $  while the other brane situated at $y = r \pi$ (where $r$ is the stable value of the 
modulus) is characterised by a negative brane tension $ {\cal V}_{vis}$ commonly called the visible brane where our Universe is located. 
The most  general metric is taken by incorporating a radion field which is a function of both spacetime co-ordinates $x^{\mu}$ and extra 
dimensional co-ordinate $y$ \cite{shiromizu_2}.
\\[2mm]
\textbf{\underline{Metric Ansatz}} :\\[1mm]
\beq
ds^2\,=\,e^{2 \phi(y,x)}dy^2\,+\,q_{\mu \nu}(y,x)dx^{\mu}dx^{\nu}    \label{2}
\eeq
The proper distance between the two branes within the fixed interval $y\,=\,0 $ to $y\,=\,r\pi$ is given by:
\beq
d_0(x)\,=\,\dis \int_{0}^{r\pi}dy \,e^{\phi{(y,x)}}     \label{36}
\eeq
The effective Einstein's equations on a $3$-brane is given by Gauss-Codacci equations as follows :
\beq
^{(4)}G^{\mu}_{\nu}\,=\,\dis \frac{3}{l^2}\,\delta^{\mu}_{\nu}\,+\,K K^{\mu}_{\nu}\,-\,K^{\mu}_{\alpha}K^{\alpha}_{\nu}\,+\,\dis \frac{1}{2}\,\delta^{\mu}_{\nu}\left(K^2\,-\,K^{\alpha}_{\beta}K^{\beta}_{\alpha}\right)\,-\,E^{\mu}_{\nu}
\eeq
and 
\beq
D_{\nu}K^{\nu}_{\mu}\,-\,D_{\mu}K\,=\,0
\eeq
where $D_{\mu}$ is the covariant derivative with respect to the induced metric $q_{\mu \nu}$ on a 
brane and $l$ is the bulk curvature radius which is related to five dimensional bulk 
cosmological constant $\Lambda_5$ as $l = \sqrt{\dis\frac{-3}{\kappa^2 \Lambda_5}}$. $K_{\mu \nu}$ is the extrinsic curvature on $y=$ constant 
hypersurface  and is given by,
\beq
K_{\mu \nu}\,=\,\nabla_{\mu}\,n_{\nu}+\,n_{\mu}D_{\nu}\phi
\eeq
where $n = e^{-\phi}\partial_{y}$ and $E^{\mu}_{\nu}$ is the projected part of the five dimensional Weyl tensor. 
In a single brane system, $ E^{\mu}_{\nu} $ gets contribution from Kaluza-Klein modes whose contribution in the low energy effective theory can be neglected \cite{shiromizu_1}. However, in the presence of two branes $ E^{\mu}_{\nu} $ does not vanish in the low energy limit due to the existence of radion fields . Therefore the evolution equations of five dimensional projected Weyl tensor and extrinsic curvature have to be solved in the bulk for determining the effective Einstein's equations. Now the junction conditions on the $3$-branes are as follows :
\beq
\left[K^{\mu}_{\nu}\,-\,\delta^{\mu}_{\nu} K\right]_{y=0}\,=\,-\dis\frac{\kappa^2}{2}\left(-{\cal V}_{pl}\, \delta^{\mu}_{\nu}\,+\,T_1^{\mu}\,_{\nu}\right)
\eeq
\beq
\left[K^{\mu}_{\nu}\,-\,\delta^{\mu}_{\nu} K\right]_{y=r \pi}\,=\,\dis\frac{\kappa^2}{2}\left(-{\cal V}_{vis}\, \delta^{\mu}_{\nu}\,+\,T_2^{\mu}\,_{\nu}\right)
\label{1}
\eeq
where $T_1^{\mu}\,_{\nu}$ and $T_2^{\mu}\,_{\nu}$ are respective energy momentum tensors on positive and negative tension branes.
\noindent In order to derive the low energy effective theory, a perturbative scheme is adopted in which the dimensionless perturbation parameter is $\epsilon = (\frac{l}{L})^2$ such that $L>>l$, where $L$ is the brane curvature scale. In order to determine $K^{\mu}_{\nu} $ and $ E^{\mu}_{\nu} $ in the bulk, these are expanded as,
\beq
K^{\mu}_{\nu}\,=\,^{(\,0\,)}K^{\mu}_{\nu}\,+\,^{(\,1\,)}K^{\mu}_{\nu}\,+....
\eeq
and
\beq
E^{\mu}_{\nu}\,=\,^{(1)}E^{\mu}_{\nu}\,+....
\eeq
\noindent At the \underline{zero-th} order we have $ ^{(0)}E^{\mu}_{\nu} = 0$ \cite{shiromizu_2, shiromizu_1, tanaka}.  There is only 
one evolution equation corresponding to $^{(0)}K^{\mu}_{\nu} $ whose solution, satisfying the Coddacci equation, is given by,
\beq
^{(0)}K^{\mu}_{\nu}\,=\,-\dis \frac{1}{l}\delta^{\mu}_{\nu}
\eeq
The junction conditions at the zero-th order are given by :
\beq
\left[^{(0)} K^{\mu}_{\nu}\,-\,\delta^{\mu}_{\nu}\, ^{(0)} K\right]_{y=0}\,=\,\dis\frac{\kappa^2}{2}\,{\cal V}_{pl}\,\delta^{\mu}_{\nu}
\eeq
and 
\beq
\left[^{(0)} K^{\mu}_{\nu}\,-\,\delta^{\mu}_{\nu}\, ^{(0)} K\right]_{y=r \pi}\,=\,-\dis\frac{\kappa^2}{2}\,{\cal V}_{vis}\,\delta^{\mu}_{\nu}
\eeq
implying that the relation between bulk curvature radius $l$ and brane tensions is 
similar to the fine-tuning condition of the RS model \cite{randall} and is given by :
\beq
\dis \frac{1}{l}\,=\,\dis \frac{1}{6}\kappa^2{\cal V}_{pl}\,=\,-\dis \frac{1}{6}\kappa^2{\cal V}_{vis}        \label{cur_rad}
\eeq
In this perturbative approach, at the zero-th order, the effect of bulk curvature and 
the brane tensions on the $3$-branes exactly counterbalance one another and as a result we have : $^{(4)}G^{\mu}_{\nu}\,=\,0$. 
This is the static RS model \cite{randall} which considers both static and flat $3$-branes. So the curvature of four dimensional 
spacetime can emerge only from higher order corrections to $K^{\mu}_{\nu}$ and $E^{\mu}_{\nu}$. \\ 
\noindent Now at the \underline{first} order we have two evolution equations, 
one for $^{(1)}E^{\mu}_{\nu}$ and the other for $^{(1)} K^{\mu}_{\nu}$ which is solved 
using the solution of evolution equation of $^{(1)}E^{\mu}_{\nu}$. The corresponding junction conditions at the first order now become :
\beq
\left[^{(1)} K^{\mu}_{\nu}\,-\,\delta^{\mu}_{\nu}\, ^{(1)} K\right]_{y=0}\,=\,-\dis\frac{\kappa^2}{2}\,T_1^{\mu}\,_{\nu}         \label{junc1}
\eeq
and 
\beq
\left[^{(1)} K^{\mu}_{\nu}\,-\,\delta^{\mu}_{\nu}\, ^{(1)} K\right]_{y=r \pi}\,=\,\dis\frac{\kappa^2}{2}\,T_2^{\mu}\,_{\nu}             \label{junc2}
\eeq
while the Gauss equation at the first order on the visible brane is given by :
\beq
^{(4)}G^{\mu}_{\nu}\,=\,-\dis \frac{2}{l} \left(^{(1)} K^{\mu}_{\nu}(y_0,x)\,-\,\delta^{\mu}_{\nu}\, ^{(1)} K(y_0,x)\right)\,-\,^{(1)}E^{\mu}_{\nu}(y_0,x) \,=\,-\dis\frac{\kappa^2}{l}\,T_2^{\mu}\,_{\nu}\,-\,^{(1)}E^{\mu}_{\nu}(y_0,x)            \label{g_1}
\eeq
where $y_0= r\pi$. Now using the bulk solution of $^{(1)}K^{\mu}_{\nu}$ which 
contains $^{(1)}E^{\mu}_{\nu}$ and eqn.{(\ref{junc1})}, we can rewrite eqn.{(\ref{junc2})} from which $^{(1)}E^{\mu}_{\nu}$ can be 
explicitly determined on the visible brane. Substituting $^{(1)}E^{\mu}_{\nu}$ in eqn.{(\ref{g_1})}, the Einstein's equation on 
visible brane is found to be :
\beq
\barr{rcl}
^{(4)}G^{\mu}_{\nu} & = & \dis\frac{\kappa^2}{l} \dis\frac{1}{\Phi}\,T_2^{\mu}\,_{\nu}\,+\,\dis\frac{\kappa^2}{l}\dis\frac{(1\,+\,\Phi)^2}{\Phi}\,T_1^{\mu}\,_{\nu}\\ [4mm]
                  & + & \dis\frac{1}{\Phi}\,(D^{\mu}D_{\nu}\Phi\,-\delta^{\mu}_{\nu} D^2 \Phi\,)\\  [4mm]
                  & + & \dis\frac{\omega(\Phi)}{\Phi^2}\left(D^{\mu} \Phi D_{\nu}\Phi\,-\,\frac{1}{2}\,\delta^{\mu}_{\nu}(D \Phi)^2\right)       \label{E1}
\earr
\eeq
where, 
\beq
\Phi\,=\,e^{2d_0(x)/l}\,-\,1, \qquad  \omega(\Phi)\,=\,-\dis \frac{3}{2} \dis \frac{\Phi}{1 \,+\,\Phi }
\eeq
So $\Phi$ is a function of the brane co-ordinates $x$. From RHS of eqn.{(\ref{E1})} we note that an effective 
brane matter on a $3$-brane can originate from :
\begin{itemize}
\item Explicit matter distribution on the brane through $T_i^{\mu}\,_{\nu}$, where $i=1,2$. 
Even a matter distribution on the hidden brane i.e. $T_1^{\mu}\,_{\nu}$ can induce a non-zero 
matter distribution on the visible brane via the bulk curvature.\\ [1mm]
and/or
\item A time dependent modulus field $e^{2 \phi(t)}$ can serve as an induced non-vanishing energy momentum tensor on the visible brane even 
in the absence of any explicit brane matter on either of the $3$-branes. 
\end{itemize}
Whether the above two scenarios may result in generating an effective cosmological constant on the visible 3-brane 
will be explored in the following sections.  \\

\section{The generalized Randall Sundrum braneworld scenario}     \label{sec_2}
We now briefly review a previous work by one of the authors which deals with non-flat brane scenario with constant radion field \cite{cosmo_con}. 
We shall subsequently follow this analysis as well as the analysis given in \cite{shiromizu_2} to explore the 
possible origin of brane cosmological constant in an expanding Universe. 
We now consider a situation in which the inter-brane distance is constant. We know that in the Randall Sundrum model (RS) \cite{randall} 
the induced cosmological constant on both the $3$-branes identically vanishes to zero and therefore to any four dimensional observer 
these branes appear flat as well as static. We can however construct a more generalized scenario \cite{cosmo_con} with a 
general warp factor that brings in curvature on both the branes and at the same time 
resolves the gauge hierarchy problem without introducing any unnatural fine tuning. Since the radion field 
is constant, therefore metric in eqn.{(\ref{2})} suggests $\phi(y,x) = 0$ and in this case the metric is taken to be,
\beq
ds^2\,=\,e^{-2A(y)}g_{\mu \nu} dx^{\mu}dx^{\nu}\,+\,dy^2       \label{3}
\eeq
The induced metric $q_{\mu \nu}(x,y)$ in the previous section is now taken as : $e^{-2A(y)}g_{\mu \nu}(x)$.
The action is :
\beq
S\,=\,\dis \int d^5x \sqrt{-G}(M^3\,{\cal R}\,-\,\Lambda_5)\,+\,\int d^4x \sqrt{-g_i}\,  {\cal V}_{i}       
\label{action1}
\eeq
where $\Lambda_5$ is AdS five dimensional bulk cosmological constant, ${\cal R}$ is the five dimensional 
Ricci scalar and $g_{\mu \nu}$ is the effective four dimensional metric. The two $3$-branes only contain the 
brane tensions ${\cal V}_{i}$ (where $i = vis, pl$) and therefore the energy-momentum tensors : $T_1^{\mu}\,_{\nu} = T_2^{\mu}\,_{\nu} = 0$. 
Now by varying the action the corresponding bulk Einstein's equations are as follows :
\beq
^{(4)}\,G_{\mu \nu}\,-\,g_{\mu \nu}e^{-2A}\,\left(-6A'^2\,+\,3A''\right)\,=\,-\dis \frac{\Lambda_5}{2 M^3}g_{\mu \nu}e^{-2A}   \label{EE1}
\eeq
and
\beq
-\dis \frac{1}{2}e^{2A} \,^{(4)}\, R\,+\,6A'^2\,=\,-\dis \frac{\Lambda_5}{2 M^3}               \label{EE2}
\eeq
with the boundary conditions :
\beq
A'(y)\,=\,\dis \frac {\epsilon_i}{12 M^3} {\cal V}_i  \qquad \epsilon_{pl}\,=\,-\epsilon_{vis}\,=\,1          \label{bc}
\eeq
\noindent where $'$ represents derivative with respect to $y$. Dividing both sides of eqn.{(\ref{EE1})} by $g_{\mu \nu}$ and after rearranging terms we 
get,
\beq
^{(4)}G_{\mu \nu}\,=\,-\Omega g_{\mu \nu}         \label{ind}
\eeq
So eqn.{(\ref{ind})} gives the effective four dimensional Einstein's equation on the Planck brane 
which is a $y=0$ hypersurface embedded in five dimensional AdS bulk. This equation being a function of $x^{\mu}$ alone, is defined on the Planck brane 
with metric $g_{\mu \nu}(x)$. The arbitrary proportionality constant  $\Omega$ therefore behaves as a induced 
cosmological constant on the Planck brane. Symbolically, we can write,
\beq
\frac{^{(4)}G^{(pl)}_{\mu \nu}}{g^{(pl)}_{\mu \nu}}\,=\,-\Omega_{pl}   \label{41}
\eeq
where
\beq
-\Omega_{pl}\,=\,e^{-2A}\left[-6A'^2\,+\,3A''\,-\,\dis \frac{\Lambda_5}{2M^3}\right]
\eeq
calculated at $y = 0$.\\
The induced metric on the visible brane is : $g^{(vis)}_{\mu \nu} = e^{-2A(\pi)}\,g^{(pl)}_{\mu \nu}$. However the Einstein's 
tensor is invariant under the multiplicative factor $e^{-2 A(\pi)}$ therefore, 
\beq
^{(4)}G^{(pl)}_{\mu \nu}\,=\, ^{(4)}G^{(vis)}_{\mu \nu}    \label{42}
\eeq
Now using eqn.{(\ref{41})} and eqn.{(\ref{42})}, the Einstein's equation $^{(4)}G^{(vis)}_{\mu \nu}$ of visible 
brane located at $y=r\pi$ orbifold fixed point is related to visible brane cosmological constant as :
\beq
\frac{^{(4)}G^{(vis)}_{\mu \nu}}{e^{-2 A(\pi)}g^{(pl)}_{\mu \nu}}\,=\,-\Omega_{pl}\,e^{2 A(\pi)}\,=\,-\Omega_{vis}
\eeq
which in turn implies that the induced cosmological constants of our Universe and that of the Planck brane are interrelated in the following way :
\beq
\Omega_{vis}\,=\,e^{2 A(\pi)}\,\Omega_{pl}     \label{43}
\eeq
Till this part of our analysis, the origin of induced cosmological constant on both the $3$-branes  is unknown. We shall however explore the 
possible origin of induced brane cosmological constant in the upcoming sections. Now using Einstein's equations one can write :
\beq
6A'^2\,=\,-\dis \frac{\Lambda_5}{2 M^3}\,+\,2\Omega_{pl} e^{2A}          \label{ind1}
\eeq
\beq
3A''\,=\,\Omega_{pl} e^{2A}
\eeq
Solution of Eqn.{(\ref{ind1})} yields a warp factor which is different from that of the original RS model. The sign of cosmological constant 
decides whether the visible brane will be de-Sitter or Anti de-Sitter.\\[1mm]
\underline{\textbf{Case-1 : $\omega^2$ is negative}}\\[1mm]
We define a dimensionless quantity on the Planck brane $ \omega_{pl}^2 = -\Omega_{pl}/3\tilde{k}^2 \geq 0 $ in 
the background of AdS bulk ($\Lambda_5 < 0$). On solving eqn.{(\ref{ind1})} the warp factor becomes,
\beq
e^{-A}\,=\,\omega_{pl} \cosh\left(\ln \dis \frac{\omega_{pl}}{c_1}\,+\,\tilde{k}|y|\right)          \label{warp1}
\eeq
where $\tilde{k} = \dis\sqrt{\frac{-\Lambda_5}{12 M^3}}$. The constant of integration $c_1\,=\,1\,+\,\sqrt{1\,-\,\omega_{pl}^2}$ is 
fixed by normalising the warp factor to unity at the orbifold point $ y = 0 $. The RS 
result of $A = \tilde{k}|y|$ is obtained in the limit $\omega_{pl} \longrightarrow 0$. It is to be 
noted that $\Omega_{pl} \longrightarrow 0$ implies $\Omega_{vis} \longrightarrow 0$  from eqn.{(\ref{43})}.
Using the boundary conditions the brane tensions on both the branes can be determined \cite{cosmo_con}. 
Now all mass scales on the visible brane get exponentially warped. 
If we define $ e^{-A(\tilde{k} r \pi)} = m/m_0 = 10^{-16} $ and $ \omega_{vis}^2 = 10^{-124} $ (say), in anti de-Sitter 
spacetime, the gauge hierarchy problem is resolved for two different values of modulus 
namely  $\tilde{k} r_1\,\simeq\,36.84\,+\,10^{-93}$ and $\tilde{k} r_2\,= 79.60$ instead of one as in the RS case.\\ [1mm]
\underline{\textbf{Case-2 : $\omega^2$ is positive}} \\[1mm]
Present cosmological observations suggest that our Universe is endowed with a positive cosmological constant i.e. has a de-Sitter character.
We therefore focus into Case-2 more critically.
By solving eqn.{(\ref{ind1})}, the warp factor is,
\beq
e^{-A}\,=\,\omega_{pl} \sinh\left(\ln \dis \frac{c_2}{\omega_{pl}}\,-\,\tilde{k}|y|\right)   \label{warp2}
\eeq
where $ \omega_{pl}^2 = \Omega_{pl}/3\tilde{k}^2 $ with $ c_2 = 1\,+\,\sqrt{1 + \omega_{pl}^2} $. 
However the induced cosmological constant on the Planck brane ($\omega_{pl}$) can be taken to be much smaller compared to $1$, so that $c_2 \approx 2$. 
For $ m/m_0 = 10^{-n}$, eqn.{(\ref{warp2})} now becomes,
\beq
e^{-\tilde{k} r \pi}\,=\,\dis \frac{10^{-n}}{2}\left[1\,+\,\sqrt{1\,+\,\omega_{pl}^2 10^{2n}}\right]         \label{warp_2}
\eeq 
A positive induced cosmological constant on the visible brane ($\omega_{vis}^2 = \omega_{pl}^2 10^{2n}$) 
does not have any upper bound and therefore can be of arbitrary magnitude. However, using the present estimated 
value of $\omega_{vis}^2 \sim 10^{-124}$ and $n=16$ we find $\tilde{k}r\pi = 36.84$ which is quite close to RS value. 
Therefore the hierarchy problem is also solved for small positive brane cosmological 
constant. Using eqn.{(\ref{bc})} the brane tensions on both the branes are given by :
\beq
{\cal V}_{vis}\,=\,-12 M^3 \tilde{k} \dis \left[\frac{c_2^2+\omega_{vis}^2}{c_2^2-\omega_{vis}^2}\right],\qquad  {\cal V}_{pl}\,=\,12 M^3 \tilde{k} \dis \left[\frac{c_2^2\,+\,{\omega_{pl}^2}}{c_2^2\,-\,{\omega_{pl}^2}}\right]         \label{V}
\eeq
where on the visible brane : $\omega^2_{vis} = \dis\frac{\Omega_{vis}}{3\tilde{k}^2} = e^{2\tilde{k}r\pi} \omega^2_{pl}$. 
But we have seen from the previous section that the contribution to non-zero value of induced 
cosmological constant $\omega_{vis}^2$ on the visible brane comes from the first order correction to the extrinsic curvature and projected Weyl tensor. 
Therefore in the zero-th order i,e. in the absence of any matter on the visible brane the RS fine-tuning condition is 
exactly valid. 
As a result in this order $\omega_{vis}^2 = 0$ and from eqn.{(\ref{V})} one finds that
${\cal V}_{vis}\,=\,-12 M^3 \tilde{k}$ and ${\cal V}_{pl}\,=\,12 M^3 \tilde{k}$. These lead to the original flat brane RS model. 
However in order to explore the origin of cosmological constant on our Universe (i,e. $\omega^2_{vis} \neq 0$ ) which is 
located at the $y= r \pi$ $3$-brane, we assume that the Universe is described by FRW metric embedded in five dimensional  AdS bulk spacetime 
containing the bulk cosmological constant $\Lambda_5$. The inter-brane separation, known as the radion field may be constant, time dependent 
or $y$-dependent. We examine each case separately to explore the possibility of obtaining 
a non-vanishing $\omega^2$ on the visible $3$-brane.\\

\section{Warped cosmological metric in constant radion field }         \label{sec_3}
We first consider a constant modulus scenario where the $3$-branes are endowed with matter densities $\rho_{vis}$, $\rho_{pl}$ and brane pressures $p_{vis}, p_{pl}$. We look for a  FRW solution on a visible brane with appropriate warp factor.\\[1mm] 
\textbf{\underline{Metric ansatz}} :
\beq
ds^2\,=\,e^{-2A(y)}\left[-dt^2\,+\,v^2(t)\delta_{ij}\,dx^{i}dx^{j}\right]\,+\,dy^2   \label{metric1}
\eeq
It is to be noted that for $\phi=0$ (i.e. for a constant radion field) and $q_{\mu \nu}(x,y) = e^{-2A(y)}g_{\mu \nu}(x)$ where $g_{\mu \nu}(x)$ is a FRW metric with a flat spatial curvature. The metric given in eqn.{(\ref{2})} reduces to that in eqn.{(\ref{metric1})}.  The location of the two $3$-branes are $y = 0, r\pi$. Since the radion field is constant therefore the proper distance along the $y$ direction between the interval $y = 0$ to $y = r \pi$ is given by:
\beq 
d_0\,=\,\int_{0}^{r \pi}dy\,=\,r\pi    \label{length1}
\eeq 
The five dimensional bulk-brane action is :
\beq
S\,=\,\dis \int d^5x \sqrt{-G}(M^3\,{\cal R}\,-\,\Lambda_5)\,+\,\dis\int d^5x \left[\sqrt{-g_{pl}}\left({\cal L}_1\,-\,{\cal V}_{pl}\right)\delta (y)\,+\, \sqrt{-g_{vis}}\left({\cal L}_2\,-\,{\cal V}_{vis}\right)\delta (y-\pi)\right]  \label{action2}
\eeq
where ${\cal L}_1$, ${\cal L}_2$ and ${\cal V}_{pl}$ , ${\cal V}_{vis}$ are Lagrangian densities and the constant brane tensions on the Planck and the visible brane respectively.
\\[1mm]
\underline{Notations}:\\  [1mm]
Non-compact brane co-ordinates : $x^{\mu}$ where $\mu\,=\,0,1,2,3$.\\
Bulk co-ordinates:  $M,N$ run over $M,N\,=\,0,1,2,3,y$.\\[1mm]
By varying the action, the five dimensional Einstein's equations can be written as :
\beq
\sqrt{-G}\,G_{M N}\,=\,-\dis\frac{\Lambda_5}{2 M^3}\,\sqrt{-G}\,g_{M N}\,+\,\frac{1}{2 M^3}\left[\tilde {T_1}^{\gamma}\,_{\mu}\,\delta(y)\sqrt{-g_{pl}}\,+\,\tilde {T_2}^{\gamma}\,_{\mu}\,\delta(y-\pi)\sqrt{-g_{vis}}\right] \, g_{\gamma \nu}\,\delta^{\mu}_M\,\delta^{\nu}_{N}     \label{EE}
\eeq
where $ g_{\gamma \nu}$ is the four-dimensional effective metric on the Planck brane. The energy momentum tensors derived from the Lagrangian densities and brane tensions on the two $3$-branes placed at the edges of the bulk are given by :\\ [1mm]
\underline{On the Planck brane} ($y=0$) :
\beq
\tilde {T_1}^{\gamma}\,_{\mu}\,=\,diag(-\rho_{pl} - {\cal V}_{pl}\,,p_{pl} - {\cal V}_{pl}\,,p_{pl} - {\cal V}_{pl}\,,p_{pl} - {\cal V}_{pl},0)
\eeq
\underline{On the visible brane} ($y=\pi$) :
\beq
\tilde {T_2}^{\gamma}\,_{\mu}\,=\,diag(-\rho_{vis}- {\cal V}_{vis}\,,p_{vis}- {\cal V}_{vis}\,,p_{vis}- {\cal V}_{vis}\,,p_{vis}- {\cal V}_{vis},0)
\eeq
On substituting eqn.{(\ref{metric1})} in eqn.{(\ref{EE})} we obtain the Einstein's equations for the above metric. Now considering the bulk part only, we have :\\[2mm]
\textit{\underline{tt component}} :\\[2mm]
\beq
\barr{rcl}
3\dis\frac{\dot{v}^2}{v^2}\,+\, e^{-2 A(y)}\,\left(\,3A''\,-\,6 A'^2\right) \,=\,-\dis\frac{\Lambda_5}{2 M^3}\,(-1)\,e^{-2 A(y)}   \label{18}
\earr          
\eeq
\textit{\underline{ii component}} :\\[2mm]
\beq
\barr{rcl}
\dis\left(-2\frac{\ddot{v}}{v}\,-\,\frac{\dot{v}^2}{v^2}\right)\,+\, e^{-2 A(y)}\left[-3A''\,+\,6 A'^2\right] \,= \,-\dis\frac{\Lambda_5}{2 M^3}\, e^{-2 A(y)}  \label{19}
          
\earr
\eeq
\textit{\underline{yy component}} :\\[2mm]
\beq
6\,A'^2\,-\,3 \,e^{2 A} \,\frac{\dot{v}^2\,+\,\ddot{v}}{v^2} \,=\,-\dis\frac{\Lambda_5}{2 M^3}   \label{20}
\eeq
where dot represents derivative with respect to time. With a little rearrangement of terms in $tt$ and $ii$ components we can write,
\beq
\dis\frac{^{(4)}G_{\mu \nu}}{g_{\mu \nu}}\,=\, e^{-2 A(y)}\,\left[-\dis\frac{\Lambda_5}{2 M^3}\,+\,3A''\,-\,6 A'^2\right]\,=\,-\Omega   \label{21}
\eeq
The left hand side is the function of time while right hand side is function of $y$ alone for every component of $g_{\mu \nu}$. This separation of variables enables us to write the effective Einstein equation on a $3$-brane as :
\beq
\dis\frac{^{(4)}G_{\mu \nu}}{g_{\mu \nu}}\,=\,-\Omega\,   
\eeq
Once again, this is the effective Einstein's equation on the Planck brane whose induced metric is $g^{(pl)}_{\mu \nu}$. Therefore the above equation may be written as,
\beq
\frac{^{(4)}G^{(pl)}_{\mu \nu}}{g^{(pl)}_{\mu \nu}}\,=\,-\Omega_{pl}   \label{23}
\eeq
such that
\beq
-\Omega_{pl}\,=\,e^{-2 A(y)}\,\left[-\dis\frac{\Lambda_5}{2 M^3}\,+\,3A''\,-\,6 A'^2\right] 
\eeq
calculated at $y=0$.\\
Here $\Omega_{pl}$ is a separation constant and is interpreted as induced cosmological constant on the Planck brane on which the warp factor is equal to unity. Since all length scales are exponentially warped on the visible brane, therefore the effective metric on the visible brane located at $y=r\pi$ hypersurface is : $g^{(vis)}_{\mu \nu} = e^{-2A(\pi)}\,g^{(pl)}_{\mu \nu}$. From the fact that Einstein's tensor remains invariant under a constant multiplicative factor, on our Universe, using Eqn.{(\ref{23})}, we have,
\beq
\frac{^{(4)}G^{(vis)}_{\mu \nu}}{e^{-2 A(\pi)}g^{(pl)}_{\mu \nu}}\,=\,-\Omega_{pl}\,e^{2 A(\pi)}\,=\,-\Omega_{vis}         \label{44} 
\eeq  
which implies
\beq
\Omega_{vis}\,=\,e^{2 A(\pi)}\,\Omega_{pl}     \label{45}
\eeq
Now in order to determine $\Omega_{vis}$ we must find the warp factor on the visible brane which will be determined very shortly. But before that let us first find the solution of the scale factor.\\
From $tt$ and $ii$ components, we have,
\beq
-\dis\frac{\dot{v}^2}{v^2}\,+\,\dis\frac{\ddot{v}}{v}\,=\,0
\eeq
On solving we get,
\beq
v(t)\,=\,e^{H_0 t}   \label{24}
\eeq
where $H_0$ is an integration constant. It is to be noted that $H_0$ is not the Hubble parameter of the effective theory which will be determined later.
Now using eqn.{(\ref{18})}, eqn.{(\ref{19})}, eqn.{(\ref{23})} and the scale factor solution given by eqn.{(\ref{24})} we get,
\beq
\Omega_{pl}\,=\,3 H_0^2    \label{26}
\eeq
which indicates a de-Sitter spacetime.  When $H_0 \longrightarrow 0$, eqn.{(\ref{26})} shows that the induced cosmological constants on both the branes vanish leading to a static and flat Universe. Now substituting eqn.{(\ref{24})} in eqn.{(\ref{20})}, the $yy$ component becomes,
\beq
A'^2\,=\,\tilde{k}^2\,+\,\omega_{pl}^2\,\tilde{k}^2\,e^{2 A}                \label{28}
\eeq
where $\tilde{k}\,=\,\sqrt{\dis\frac{-\Lambda_5}{12 M^3}}$ and $\omega_{pl}^2\,=\,\dis\frac{\Omega_{pl}}{3\tilde{k}^2}$ is a dimensionless quantity. Similarly one can also define $\omega_{vis}$ on the visible brane such that : $\omega_{vis}^2\,=\,\dis\frac{\Omega_{vis}}{3\tilde{k}^2}= \dis\frac{e^{2 A(\pi)}\Omega_{pl}}{3\tilde{k}^2}$.  Now on solving eqn.{(\ref{28})}, the solution of the warp factor consistent with $Z_2$ symmetry is given by :
\beq
e^{-A(y)}\,=\,\omega_{pl}\,\sinh\left[-\tilde{k}\,|y|\,+\,\ln\dis\frac{c_2}{\omega_{pl}}\right]   \label{25}
\eeq
where $c_2$ is the integration constant. We normalize the warp factor on the Planck brane, such that $e^{-A(y = 0)}\,=\, 1$ so that eqn.{(\ref{25})} can be written as,
\beq
\omega_{pl}\,\sinh\left[\ln\dis\frac{c_2}{\omega_{pl}}\right]\,=\,1
\eeq
After a little rearrangement of terms we finally get :
\beq
c_2\,=\,1\,+\,\sqrt{1\,+\,\omega^2_{pl}}  \label{29}
\eeq
Taking the value of $\omega^2_{pl}$ much less compared to $1$, $c_2 \approx 2$.\\
Now substituting eqn.{(\ref{24})} and eqn.{(\ref{25})} with $c_2\,=\,2$ in eqn.{(\ref{metric1})}, the resulting $5$D metric becomes :
\beq
ds^2\,=\,\omega_{pl}^2\,\sinh^2 \left[-\tilde{k} \,|y|\,+\,\ln\dis\frac{2}{\omega_{pl}}\right]\left(-dt^2\,+\,e^{2 H_0 t}\delta_{ij}dx^{i}\,dx^{j}\right)\,+\,dy^2      \label{10}
\eeq
The above metric describes exponential expansion in the spatial three dimensions. In the static limit,  $\omega_{pl} \longrightarrow 0$ and $\omega_{vis} \longrightarrow 0$ that imply $H_0$ also becomes zero. Eqn.{(\ref{10})} now corresponds to the static RS metric \cite{randall}. Since the radion is independent of spacetime as well as extra-dimensional co-ordinates, all terms involving covariant derivatives in eqn.{(\ref{E1})} vanish. Therefore the effective Einstein's equations on the visible brane are only related to energy momentum tensors of the two $3$-branes and bulk curvature radius $l$: 
\beq
^{(4)}G^{\mu}_{\nu} \,= \, \dis\frac{\kappa^2}{l} \dis\frac{1}{\Phi}\,T_2^{\mu}\,_{\nu}\,+\,\dis\frac{\kappa^2}{l}\dis\frac{(1\,+\,\Phi)^2}{\Phi}\,T_1^{\mu}\,_{\nu}    \label{eff_EE}
\eeq
where $\Phi\,=\,(e^{2 r \pi/l}\,-\,1) $, $\kappa^2$ is related to the five dimensional gravitational constant. Now from eqn.{(\ref{44})} and eqn.{(\ref{eff_EE})} the visible brane cosmological constant is :
\beq
-\Omega_{vis}\,=\,\dis\frac{\kappa^2}{4 l (e^{2 r\pi/l}-1)}\,\left[e^{4 r \pi/l}\,T_1^{\mu}\,_{\nu}\,\delta^{\nu}_{\mu}\,+\,T_2^{\mu}\,_{\nu}\,\delta^{\nu}_{\mu}\right]    \label{w_vis}
\eeq
So the existence of induced cosmological constant on the visible brane is directly related to the presence of matter density and pressure of both the $3$-branes. The effects of extra dimension on the induced brane cosmological constant however shows up through the multiplicative factor $\dis\frac{\kappa^2}{4 l (e^{2 r \pi/l}-1)}$.  Furthermore it is interesting to find that even if the visible brane matter $T_2^{\mu}\,_{\nu} = 0$, there can be a net cosmological constant in the Universe solely due to the matter content of the hidden brane. We now determine the effective Hubble parameter of our Universe.  
\subsection{Effective Hubble parameter on the visible brane :}
Now, we determine the Hubble parameter on the $4$D world following the method adopted in \cite{kim}. The induced metric of the four dimensional spacetime is :
\beq
ds_{(4)}^2 = -d\tilde{t}^2\,+\,e^{2 H(y)\tilde{t}}\delta_{ij}\,d\tilde{x}^i\,d\tilde{x}^j   \label{46}
\eeq
Here, the effective Hubble parameter is determined on orbifold fixed points i.e. at $y = 0, r\pi$. In                                                                                                                                                                                                                                                                                                                                                                                                                                                                        order to get the induced metric from $5$D metric, we define following co-ordinate transformations :
\beq
d\tilde{t}\,=\,\omega_{pl}\sinh\left[-\tilde{k}\,|y|\,+\,\ln\dis\frac{2}{\omega_{pl}}\right]\,dt        \label{27}
\eeq
and 
\beq
d\tilde{x}^i\,=\,\omega_{pl}\sinh\left[-\tilde{k}\,|y|\,+\,\ln\dis\frac{2}{\omega_{pl}}\right]\,dx^i
\eeq
Now for a fixed $y$ (on a $3$-brane) integrating eqn.({\ref{27}}) results into :
\beq
\tilde{t}\,=\,\omega_{pl}\sinh\left[-\tilde{k}\,|y|\,+\,\ln\dis\frac{2}{\omega_{pl}}\right]\,t
\eeq
On comparing the $5$D metric and $4$D effective metric given by eqn.({\ref{10}}) and eqn.({\ref{46}}) we obtain,
\beq
\omega_{pl}^2 \sinh^2\left[-\tilde{k}\,|y|\,+\,\ln\dis\frac{2}{\omega_{pl}}\right]\,e^{2H_0 t}\,\delta_{ij}\,dx^i\,dx^j\,=\,e^{2\,H(y)\tilde{t}}\delta_{ij}\,d\tilde{x}^i\,d\tilde{x}^j
\eeq
Once again, if we determine the Hubble parameter on a $3$-brane by comparing the terms in the exponential, $H$ should be a function of $y$ because $t$ depends on $\tilde{t}$ through $y$.
Therefore the effective $4$D Hubble parameter on a $3$-brane for a given value of $y$ is given by :
\beq
H(y)\,=\,\dis\frac{H_0}{\omega_{pl}}\,cosech\left[-\tilde{k}\,|y|\,+\,\ln\dis\frac{2}{\omega_{pl}}\right]
\eeq
\underline{On the visible brane} ($y\,=\,r \pi$) :
\beq
H(\pi)\equiv H_{vis}\,=\,\dis\frac{H_0}{\omega_{pl}\,\sinh\left[-\tilde{k} r \pi\,+\,\ln\dis\frac{2}{\omega_{pl}}\right]}
\eeq
Since $\omega_{pl}=\dis\frac{H_0}{\tilde{k}}$ and $\omega_{vis} = \omega_{pl} \,e^{A(\pi)}$, therefore the effective visible brane Hubble parameter becomes,
\beq
H_{vis}\,=\,\tilde{k}\,cosech\left[-\tilde{k} r \pi\,+\,\ln\dis\frac{2}{\omega_{pl}}\right]\,=\,\omega_{vis}\tilde{k}  \label{H_vis}
\eeq
The above equation suggests that $H_{vis} \simeq \omega_{vis}$ (in Planckian unit). This result is consistent with the present result relating the Hubble parameter and cosmological constant of the Universe. Thus when the brane separation is constant, present evolution of the Universe is indeed governed by constant vacuum energy densities on the $3$-branes. As expected, a vanishing $H_{vis}$ implies a static as well as flat Universe, devoid of any matter. Such a Universe does not undergo exponential expansion but is described by static RS model where both the branes are flat possessing brane tensions only \cite{randall}.\\[1mm]
\noindent \underline{\textbf{Energy-momentum tensor on $3$-branes}} :\\ [2mm]
Visible brane:  $T_2^{\mu}\,_{\nu}\,=\,diag(-\rho_{vis},p_{vis},p_{vis},p_{vis})$ \\[2mm]
Hidden brane:  $T_1^{\mu}\,_{\nu}\,=\,diag(-\rho_{pl},p_{pl},p_{pl},p_{pl})$\\  [2mm]
Substituting the components of $T_i\,^{\mu}\,_{\nu}$ (where $i = 1, 2$) in eqn.{(\ref{w_vis})}, the visible brane induced cosmological constant is :
\beq
-\Omega_{vis}\,=\,\dis\frac{\kappa^2}{4\,l (e^{2 r \pi/l}-1)}\,\left[-\left(e^{4 r \pi/l}\,\rho_{pl}\,+\,\rho_{vis}\right)\,+\,3\left(e^{4 r \pi/l}\,p_{pl}\,+\,p_{vis}\right)\right]  \label{17}
\eeq
In case of vacuum energy dominated Universe, the energy density and pressure are related as :
\beq
\rho_{vis}\,=\,-p_{vis}
\eeq
Similarly, if the same equation of state is valid on Planck brane, then from eqn.({\ref{17}}) the induced cosmological constant on the visible brane becomes :
\beq
\Omega_{vis}\,=\,\dis\frac{\kappa^2}{l (e^{2 r\pi/l}-1)}\,\left[e^{4 r \pi/l}\,\rho_{pl}\,+\,\rho_{vis}\right]     \label{37}
\eeq
From the above equation we find that the induced $\Omega_{vis}$ is constant only when the visible brane contains matter 
in the form of constant vacuum energy density. Most importantly the non-zero value of brane matter 
as well as the bulk cosmological constant ($1/l$) are essential to inject an effective cosmological 
constant on the visible brane. Thus in a braneworld consisting of two $3$-branes one cannot generate an 
effective $\Omega_{vis}$ in the $4$D spacetime only out of the intrinsic brane tensions and 
bulk cosmological constant without any brane matter when modulus field is independent of spacetime co-ordinates. 
If we further focus on eqn.({\ref{37}}), we discover that an absence of matter in the visible brane i,e. $\rho_{vis} = 0$ does not 
necessarily imply a vanishing $4$D cosmological constant as long as $\rho_{pl}\neq 0$. This can be seen from eqn.({\ref{37}}) by putting $\rho_{vis} = 0$,
\beq
\Omega_{vis}\,=\,\dis\frac{\kappa^2}{l (e^{2 r \pi/l}-1)}\,{e^{4 r \pi/l}\,\rho_{pl}} \label{38}
\eeq
So from eqn.({\ref{38}}), we can infer that the Planck brane matter $\rho_{pl}$ mediates an effective  cosmological constant on the 
visible brane due to the curvature in bulk spacetime and  therefore the vacuum energy density of the Planck brane and  
five dimensional bulk cosmological constant $\Lambda_5$ may be the possible origins of an effective  cosmological constant on our 
Universe (i,e. the visible brane) which can result into an exponentially expanding Universe.\\
  
\section{Extra dimension dependent radion field}      \label{sec_4}
Let us now assume that the inter-brane distance depends only on the extra dimension $y$ i.e. $\phi$ in eqn.({\ref{2}}) is a function of $y$ only. 
This modifies the proper distance between the two branes as defined in eqn.({\ref{36}}). we explore the possibility of having  an effective 
cosmological constant on our Universe ( visible 3-brane ) placed at $y=r \pi$. In such a scenario, we consider the metric,
\textbf{\underline{Metric ansatz}} :\\[2mm]
\beq
ds^2\,=\,e^{-2A(y)}\left[-dt^2\,+\,v^2(t)\delta_{ij}\,dx^{i}dx^{j}\right]\,+\,e^{2 \phi(y)}dy^2 \label{35}
\eeq          
The proper distance along the $y$ co-ordinate between the orbifold fixed points $y = 0$ to $y = r \pi$ is now given by:
\beq 
d_0\,=\,\int_{0}^{r \pi}dy\, e^{\phi(y)}    \label{length}
\eeq
Substituting eqn.({\ref{35}}) in eqn.({\ref{EE}}), Einstein's equations in the bulk are :\\[1mm]
\textit{\underline{tt component}} :\\[2mm]
\beq
\barr{rcl}
3\dis\frac{\dot{v}^2}{v^2} \,+\, e^{-2 A}\,e^{-2 \phi}\left(3A''\,-\,6 A'^2\,-\,3 A'\phi'\right)\,=\,-\dis\frac{\Lambda_5}{2 M^3}\,(-1)\,e^{-2 A}    \label{8}
\earr          
\eeq
\textit{\underline{ii component}} :\\[2mm]
\beq
\barr{rcl}
-\dis\left(2\ddot{v}v\,+\,\dot{v}^2\right)\,+ \,  e^{-2 A}\,e^{-2 \phi}v^2\left(-3A''\,+\,6 A'^2\,+\,3\,A' \phi^{'}\right)\,=\, -\dis\frac{\Lambda_5}{2 M^3}\,e^{-2 A}v^2  
\label{ii1}          
\earr
\eeq
\textit{\underline{yy component}} :\\[2mm]
\beq
6\,A'^2\,-\,3 \,e^{2 (A\,+\,\phi)} \,\left[\dis\frac{\dot{v}^2}{v^2}\,+\,\dis\frac{\ddot{v}}{v}\right] \,=\,-\dis\frac{\Lambda_5}{2 M^3}\,e^{2 \phi}   \label{yy1}
\eeq
From $tt$ and $ii$ components, the solution of the scale factor is found to be :
\beq
v(t)\,=\,e^{H_0 t}   \label{30}
\eeq
where $H_0$ is a constant. Dividing eqn.{(\ref{8})} by $g_{tt}$,   eqn.{(\ref{ii1})} by $g_{ii}$ and using the solution of the scale factor, 
we obtain the effective Einstein's equation on the Planck brane as,
\beq
\dis\frac{^{(4)}G_{\mu \nu}^{(pl)}}{g_{\mu \nu}^{(pl)}}\,= \,\left[\dis\frac{-\Lambda_5}{2 M^3}\,e^{-2 A}\,+\,\left(-6 A'^2\,-\,3 A'\phi'\,+\,3 A''\right)\,e^{-2 (A + \phi)}\right]\,=\,-\Omega_{pl}\,=\,-3 H_0^2 \label{7}
\eeq         
However, the induced cosmological constant on the visible brane is related to the Planck brane as $\Omega_{vis} = \Omega_{pl}\,e^{2 A(\pi)}$. Since the covariant derivatives are evaluated on $y=r\pi$ hypersurface therefore using eqn.{(\ref{E1})} the effective cosmological constant on the visible brane is :
\beq
-\Omega_{vis}\,=\,\dis\frac{\kappa^2}{4 l (e^{2 d_0/l}-1)}\,\left[e^{4 d_0/l}\,T_1^{\mu}\,_{\nu}\,\delta^{\nu}_{\mu}\,+\,T_2^{\mu}\,_{\nu}\,\delta^{\nu}_{\mu}\right]   \label{12}
\eeq
where $d_0$ is constant. Therefore {(\ref{30})} and {(\ref{12})} once again suggest that even a y-dependent radion field cannot produce an exponential expansion along spatial three dimensions unless a constant energy density is given explicitly on $3$-branes. The bulk effect however appears through the bulk curvature radius $l$. It may be further noted that even when visible brane matter $T_2^{\mu}\,_{\nu} = 0$, the Planck brane matter  and $\Lambda_5$ together can induce visible brane cosmological constant $\Omega_{vis}$ that triggers an exponentially expanding Universe. However $\Omega_{vis}$ differs from {(\ref{w_vis})} in terms of proper length as defined in eqn.{(\ref{length})}.\\
\section{Radion driven acceleration without brane matter - time dependent radion field}   \label{sec_5}
So far we were studying an exponentially expanding model in the presence of brane (visible and/or hidden) matter leading to an effective $\Omega_{vis}$ on the visible brane modulated by bulk effects. In this section we show that if the radion field is time dependent then there can be an effective acceleration of the Universe even in the absence of any explicit brane matter. In this case the dynamics of the radion field produces the effective acceleration of $3$-brane scale factor. For this we recall the original RS model with a time dependent radion field and look for a possible cosmological solution in the effective $4$D theory. Following \cite{Rad_stab} we assume that in presence of a time-varying field the metric has a form:\\[1mm]
\beq
ds^2\,=\,e^{-2 k b(t)|y|}\left[-dt^2\,+\,v^2(t)\delta_{ij}\,dx^{i}dx^{j}\right]\,+\,b^2(t)\,dy^2  \label{metric3}
\eeq 
where $b(t) = e^{2 \phi(t)}$.
 In this case the proper distance between the interval $y=0$ to $y=r \pi$ is given by : \\[1mm]
\beq 
d_0(t)\,=\,\int_{0}^{r\pi}\,e^{2 \phi(t)}dy\,=\,r \pi e^{2 \phi(t)}   \label{length2}
\eeq
Such a metric is considered in \cite{Rad_stab} where the warp factor is : $e^{-k b(t)|y|}$, $b(t)$ is the radion field.In the static limit, 
when the inter-brane separation is a constant, 
this warp factor reduces to that of the warp factor of RS metric \cite{randall} where both the $3$-branes are flat and static. However 
it is to be noted that in general this warp factor does not lead to Einstein's equations separable solely in terms of $t$ and $y$ variables. 
In order to study the time evolution of our Universe located on the visible brane, we therefore consider the effective $4$D metric in 
the an appropriate choice of co-ordinates as : $ds^2_{(4)}\,=\, -dt^2\,+\,v^2(t)\,\delta_{ij}\, dx^{i} dx^{j}$, where $v(t) = v(t,\pi)$.\\[2mm]
With the above effective metric, the components of effective Einstein's equations on the visible brane together with eqn.{(\ref{E1})} become :\\[1mm]
\textit{\underline{tt component}} :\\[2mm]   
\beq
^{(4)}G^{t}_{t}\,=\,-\frac{3\dot{v}^2}{v^2}\,=\,\frac{\kappa^2}{l (e^{2 d_0/l}-1)}\,\left[T_2\,^{t}\,_{t}\,+\,e^{4 d_0/l}\,T_1\,^{t}\,_{t}\right]\,+\,\frac{3\, e^{2 d_0/l} }{l^2 (e^{2 d_0/l}-1)} \,\dot{d_0}^2   \label{31}
\eeq
\textit{\underline{ii component}} :\\[2mm]
\beq
^{(4)}G^{i}_{i}= -\dis\left(2\frac{\ddot{v}}{v} + \frac{\dot{v}^2}{v^2}\right)= \frac{\kappa^2}{l (e^{2 d_0/l}-1)} \left[T_2\,^{i}\,_{i} + e^{4 d_0/l}\,T_1\,^{i}\,_{i}\right]\,+\,\frac{2\, e^{2 d_0/l} }{l (e^{2 d_0/l}-1)} \dis\left(\frac{3\,\dot{v}}{v}\dot{d_0} + \ddot{d_0}\right) +\, \frac{e^{2 d_0/l} }{l^2 (e^{2 d_0/l}-1)} \,\dot{d_0}^2   \label{32}
\eeq
\\
where $d_0(t)$ is given by eqn.{(\ref{length2})} and the spatial component of the covariant derivative $D_{i}(e^{2 d_0(t)/l}-1) = 0$ on the visible brane. Now subtracting eqn.{(\ref{32})} from eqn.{(\ref{31})}, we get,
\beq
\barr{rcl}
\dis\frac{\ddot{v}}{v}\, -\, \dis\frac{\dot{v}^2}{v^2}\,=\,\frac{d}{dt}H(t)  & = &\dis\frac{\kappa^2}{2 \,l (e^{2 d_0/l}-1)}\,\left[(T_2\,^{t}\,_{t} - T_2\,^{i}\,_{i}) \,+\,e^{4 d_0/l}\,(T_1\,^{t}\,_{t} - T_1\,^{i}\,_{i}) \right] \\[4mm]
    & -&  \dis\frac{\, e^{2 d_0/l} }{l (e^{2 d_0/l}-1)}\,\dis\left(3\,\frac{\dot{v}}{v}\,\dot{d_0}\,+\,\ddot{d_0}\right)
    \, +  \, \dis\frac{e^{2 d_0/l} }{2\,l^2\, (e^{2 d_0/l}-1)} \,\dot{d_0}^2    \label{33}
\earr
\eeq
where $H(t) = \dis\frac{\dot{v}(t)}{v(t)}$ is the Hubble parameter of the Universe located on the visible brane. 
Eqn.{(\ref{33})} reveals that a time dependent modulus field can itself produce dynamical evolution of the Universe even in the absence of matter 
on the $3$-branes i,e. when $T_1\,^{\mu}\,_{\nu} = T_2\,^{\mu}\,_{\nu} = 0$. 
The signature of bulk in the evolution is evident from the presence of bulk curvature radius $l$. Now in case of a constant radion field 
when all terms involving $\dot{d_0},\ddot{d_0}$ drop out, from eqn.{(\ref{33})} we retrieve the exponential solution of the scale 
factor $v(t)$ on the visible brane as obtained earlier when both the $3$-brane energy-momentum tensors are described by vacuum energy densities 
for which the equation of state is $\rho_{i}= -p_{i}$ where $i = 1,2$. \\
Let us now assume a slowly time varying radion field. Then in absence of any brane matter i.e. 
$T_1\,^{\mu}\,_{\nu} = T_2\,^{\mu}\,_{\nu} = 0$, eqn.{(\ref{33})} in the leading order of $\dot{d_0}$ becomes :
\beq
\dis\frac{\ddot{v}}{v}\, -\, \dis\frac{\dot{v}^2}{v^2}\,=\,\frac{d}{dt}H(t)\,=\,\frac{3\,e^{2 d_0/l}\,\dot{d_0}}{l (e^{2 d_0/l}-1)}\,H(t)
\eeq 
which implies
\beq
\frac{d}{dt} \ln H(t)\,=\,\frac{d}{dt} \left[ \frac{3}{2}\,\ln (e^{2 d_0/l}-1)\right]
\eeq
After integration we get,
\beq
H(t)\,=\,(e^{2 d_0(t)/l}-1)^{3/2}      \label{39}   
\eeq
Since $\dot{H(t)} > 0$, therefore an accelerating phase of our Universe is obtained from a slowly time-varying modulus field. 
Thus a time-varying radion field indeed leads to an accelerating phase of visible $3$-brane even 
when the energy momentum tensors vanish on both the branes. In the presence of appropriate modulus potential, such a scenario 
can be envisioned as the tunnelling of the modulus from a metastable vacuum state to a stable vacuum state 
leading to an accelerating phase of the Universe located on the $3$-brane.

\section{Conclusion}
In the background of the work by Shiromizu et.al \cite{shiromizu_2} and generalised Randall Sundram  
braneworld scenario \cite{cosmo_con}, our focus in the present work is to study the role of brane energy momentum 
tensors $T_i\,^{\mu}\,_{\nu}$,  the bulk curvature and the modulus field in inducing an effective 
four dimensional cosmological constant on the visible $3$-brane embedded in a five dimensional AdS ($\Lambda_5 < 0$) bulk. 
We have considered three different cases to study the dynamical evolution of our Universe. Our results can be summarised as follows :
\begin{enumerate}
\item \underline{Constant modulus field}
\begin{itemize}
\item In this case an effective cosmological constant $\Omega$ appears on the visible $3$-brane due to the combined effects of the bulk curvature
and brane matters in either of the two 3-branes. On solving  Einstein's equations, we obtain an exponential solution of the 
scale factor $v(t) = e^{H_0 t}$ and $\Omega_{vis} = 3 H_0^2 \,e^{2 A(\pi)} $ which indicates four dimensional de-Sitter spacetime. 
In the static limit  $H_0\longrightarrow 0$ and therefore $\Omega_{vis} \longrightarrow 0$. This leads to static RS model with flat and static 
visible 3-brane \cite{randall}. The same argument applies for hidden brane also.
\item The nature of the warp factor is determined. It is found that the sine hyperbolic solution of the warp factor 
depends on $\Omega_{pl}$, modulus length and bulk cosmological constant. 
The induced $\Omega_{vis}$ on the visible brane is further related to energy momentum 
tensors (see eqn.{(\ref{w_vis})}) of both the $3$-branes. 
Therefore, absence of matter on both the $3$-branes corresponds to zero value of effective $4$D cosmological constant and hence 
reproduces RS warp factor \cite{randall} which signifies a static and flat visible 3-brane.
\item On the visible brane, from eqn.{(\ref{37})} it is seen that the effective brane  cosmological constant $\Omega_{vis}$ 
depends on energy momentum tensors of both the $3$-branes through constant $\rho_{vis}$ and $\rho_{pl}$. 
The effects of bulk curvature and the proper length given by $d_0 = r\pi$ on $\Omega_{vis}$ appear through the 
factor $\dis\frac{\kappa^2}{4 l (e^{2r\pi/l}-1)}$. Therefore in order to generate a non-zero $\Omega_{vis}$, the contribution of 
non-zero $T_i\,^{\mu}\,_{\nu}$ (where $i = 1,2$) as well as bulk cosmological constant $\Lambda_5$ are essential which 
can be seen from eqn.{(\ref{37})}.
\item  Furthermore eqn.{(\ref{38})} suggests that one can have a scenario where $\Omega_{vis} \neq 0$, even if our Universe is 
devoid of any matter i,e. when $\rho_{vis} = 0$ but $\rho_{pl} \neq 0$. Thus the exponential expansion of our Universe may be due to 
the presence of bulk cosmological constant and Planck brane energy density $\rho_{pl}$.
\item We have determined the Hubble parameter of the Universe for a constant radion field. 
From eqn.{(\ref{H_vis})}, $H_{vis}$ is found to depend on bulk cosmological constant $\Lambda_5$, $\Omega_{vis}$ and constant brane 
separation distance $r$. This implies that even if there is no matter on the visible brane, 
one can still obtain a non-zero Hubble parameter on the Universe with $\rho_{pl} \neq 0$ as $\Omega_{vis} \neq 0$. 
The value of Hubble parameter of our Universe from eqn.{(\ref{H_vis})} is found to be proportional to $\omega_{vis}$  of the visible brane which
matches very well with the present cosmological results.
\end{itemize}
\item\underline{{Extra dimensional co-ordinate dependent modulus field}}
\begin{itemize}
\item In case the modulus field is $y$ dependent, the expression of proper length changes 
as shown in {(\ref{length})}. In this situation, we can generate a four dimensional effective cosmological constant on the visible brane and 
hence in our Universe. This results into the scale factor which varies exponentially with time as given by eqn.{(\ref{30})}. 
Such a scale factor indicates exponential expansion of the Universe in the presence of extra dimensional co-ordinate dependent modulus 
field which guarantees four dimensional de Sitter spacetime.
\item However, such a radion field cannot produce a dynamical evolution by itself unless there are non-zero contributions from
energy momentum tensors on the $3$-branes and bulk cosmological constant $\Lambda_5$. It may further be noted that, just as in the previous case,  
an effective  $4$D cosmological constant appears on the visible brane due to a non-vanishing Planck brane matter 
even if the visible brane matter $T_2\,^{\mu}\,_{\nu} = 0$ (see {(\ref{12})}).
\end{itemize}
\item{\underline{Time-varying modulus field}}
\begin{itemize}
\item An effective energy-momentum density is induced on our Universe in a situation when the inter-brane separation varies with time.
\item The dynamical evolution of the Universe is now possible in the absence of any matter 
on the $3$-branes i,e. when $T_1\,^{\mu}\,_{\nu} = T_2\,^{\mu}\,_{\nu} = 0$. Actually this time dependent modulus field and 
bulk cosmological constant together trigger the time evolution of the 
Universe leading to a non-zero Hubble parameter.
\item In case of a slowly time varying radion field, the Hubble parameter on our Universe 
has been determined in {(\ref{39})} purely in terms of time varying proper length $d_0(t)$ (defined in {(\ref{length2})}) modulated by 
bulk curvature $l$. The fact that $\dot{H(t)} > 0$ further suggests an accelerating nature of the Universe driven 
solely by time dependent radion field.
\item Furthermore our result reveals that a solely cosmological constant driven expansion is possible only when either of the brane energy-momentum 
tensors is non-zero and the radion field is time independent i.e. a time dependent radion field can not give rise to a constant 3-brane vacuum energy. 
This can be easily seen from eqn.{(\ref{33})}.
\end{itemize}
\end{enumerate}

\end{document}